\documentclass{caosp309}
\usepackage{threeparttable}

\usepackage{graphicx}
\usepackage{float}
\usepackage{tablefootnote}
\usepackage{subcaption}

\usepackage{float}        
\usepackage{siunitx}      
\usepackage{booktabs}     
\usepackage{xcolor}       
\definecolor{darkblue}{RGB}{0,0,139}

\usepackage{threeparttable}
\usepackage{hyperref}
\usepackage{url}
\usepackage{doi}

\usepackage[authoryear]{natbib}
\articleNo{BP08}
\pubyear{2024}
\volume{54}
\volnumber{1}
\firstpage{1}
\received{February 11, 2026}
\accepted{April 22, 2026}

\begin{document}

\hauthor{P.\,Rutov\'{a}, M.\,Skarka, J. \v{Z}\'{a}k, and P.\,Gajdo\v{s}}

\title{Rotational characteristics of five photometrically variable stars}

\author{
        P.\,Rutov\'{a}\inst{1}
      \and
        M.\,Skarka\inst{2}\orcid{0000-0002-7602-0046}  
      \and
       J. \v{Z}\'{a}k\inst{2}\orcid{0000-0001-9416-9007}
       \and
        P.\,Gajdo\v{s}\inst{3, 2}\orcid{0000-0003-1478-3256} 
       }

\institute{
           Department of Theoretical Physics and Astrophysics, Faculty of Science, Masaryk University, Kotl\'a\v{r}sk\'a 267/2, 611~37 Brno, Czech Republic, \email{534281@mail.muni.cz}
         \and 
         \ondrejov
         \and
         Institute of Physics, Faculty of Science, Pavol Jozef \v{S}af\'arik University, 04001 Ko\v{s}ice, Slovakia
          }

\date{November 3, 2023}

\maketitle

\begin{abstract}
This study explores photometric variability in a sample of hot stars in order to test whether variability arises from rotational modulation due to surface temperature spots. Frequencies determined from the projected rotational velocities were compared with frequencies estimated on the basis of TESS light curves. In all five cases, the spectroscopic frequency was lower than the photometric one preserving possibility that the observed brightness variations can be rotationally induced. 
In two of our targets, HT\,Cet and HD\,219487, the photometric frequency spectra suggest more complex variability, likely pulsations. 
Our results suggest that spot activity inducing rotational modulation may exist in stars hotter than 6500\,K, highlighting the value of combined spectroscopic and photometric analysis of variations.
\end{abstract}

\section{Introduction}\label{intr}

Photometric variability in hot stars is not yet fully understood, in contrast to solar-like stars where star spots and rotation are well-established causes of quasi-periodic variations \citep[e.g.][]{Reinhold2013,McQuillan2014,Santos2021}. In hotter stars, variability may result from surface inhomogeneities and rotation, effects in close binaries, and pulsations \citep[e.g.][]{Skarka2022,Skarka2024}. Surface inhomogeneities can arise from two different mechanisms. In chemically peculiar stars, abundance spots produced by diffusion in strong magnetic fields can result in highly stable light curves with typically low photometric amplitudes \citep[e.g.][]{Kochukhov2017}. In contrast, temperature spots associated with magnetic activity, similar to starspots on the Sun, can produce rotational modulation and evolve on observable timescales \citep[e.g.][]{Berdyugina2005}. In this work, we consider the possibility that temperature spots produce rotational modulation. However, the relative roles of these factors are unclear. The objective of our study is to determine whether rotational modulation can offer an explanation for the observed photometric variability in a sample of stars hotter than 6500\,K, where this kind of stellar variability is not expected but also not ruled out \citep{Cantiello2019,Balona2022,Kraft1967}. This goal is achieved by measuring projected rotational velocities ($v\sin i$) estimated from spectra and comparing them with frequencies estimated from the data series gathered by the \textit{Transiting Exoplanet Survey Satellite} \citep[\textit{TESS,}][]{Ricker2015}.

\section{Observations}

Candidate stars were selected from the TESS input Catalogue \citep[TIC v8.2,][]{Paegert2021}, based on light curves that exhibit signs of potential rotational variability identified by \citet{Skarka2022}. 
The selection criteria included apparent brightness (Johnson $V$ magnitude $\leq$ 9), good visibility from the Ondřejov Observatory (Czech Republic), and effective temperatures above 6500\,K. 
We selected five targets for this purpose: HD\,197039, HD\,204485, HD\,219487, 5\,Peg, and HT\,Cet (see Table \ref{Tab:Parameters} with the basic parameters from the TIC catalogue).

\begin{table}
\caption{Basic parameters of the sample stars from TIC. Values of $v_{\rm mic}$ and $v_{\rm mac}$ were calculated from empirical relations (see the text for further information).}       
\label{Tab:Parameters}      
\centering     
\begin{tabular}{l | c c c c c}       
\hline\hline   
Star ID      & $T_{\rm eff}$ (K) & $\log g$ & $R$ (R$_{\odot}$) & $v_{\rm mic}$ (km/s) & $v_{\rm mac}$ (km/s) \\
\hline
HD 197039 & 6533 & 3.91 & 2.146 & 1.48 & 9.50 \\
HD 204485 & 7186 & 4.13 & 1.801 & 1.88 & 5.50 \\
HD 219487 & 6869 & 4.33 & 1.308 & 1.63 & 12.99\\
5 Peg     & 7112 & 3.37 & 4.306 & 2.77 & 3.85 \\
HT Cet    & 6929 & 4.00 & 2.020 & 1.70 & 14.06\\
\hline \hline
\end{tabular}
\end{table}

\begin{table}
\caption{Determined parameters from the analysis. Spectroscopic ($f_{\rm sp}$) and photometric ($f_{\rm ph}$) frequencies in the first two columns are in cycles per day, velocities in km/s. Column $(v\sin i)_{\rm lit}$ and 'Ref' give the literature values and the corresponding references, respectively. Inclination angles ($i$ (deg)) were derived in this work from a comparison of spectroscopic and photometric frequencies (see Section \ref{Sect:Discussion}).The errors are in the parentheses.}       
\label{Tab:vsini}      
\centering
\begin{threeparttable}
\begin{tabular}{l | c c c c c c}       
\hline\hline   
Star ID      & $f_{\rm sp}$ & $f_{\rm ph}$ & $v\sin i$ & $(v\sin i)_{\rm lit}$ & $i$ (deg) & Ref \\
\hline
HD 197039 & 0.25(3) & 0.236(4) &  27(3) &  26 & 90(-) & 1 \\
HD 204485 & 0.10(2) & 0.146(4) &  9(2)  &  9.3 & 43(12) & 2 \\
HD 219487 & 0.40(6) & 0.581(4) &  28(4) &  22-35 & 44(8)& 1\\
5 Peg     & 0.73(4) & 0.828(4) &  159(5)& 155, 170  & 62(6) & 3,4\\
HT Cet    & 0.28(5) & 0.912(4) &  29(5) & 28 & 18(3) & 1\\
\hline \hline
\end{tabular}
\begin{tablenotes}
\item[a] References: 1 -- \citet{Glebocki2005}, 2 -- \citet{Luck2017}, 3 -- \citet{Abt1995}, 4 -- \citet{Royer2002}.
\end{tablenotes}
\end{threeparttable}
\end{table}

We obtained high-resolution spectroscopic data using the Ondřejov Echelle Spectrograph \citep[OES, resolving power $R=\Delta \lambda/\lambda=50000$ at H$\alpha$,][]{Koubsky2004,Kabath2020}. In 2024-2025, we obtained one spectrum for each target reaching signal-to-noise ratio between 70 and 120 (measured at 660\,nm). The SNR of each spectrum is shown in Fig.~\ref{Fig:Spectra}. All the steps of data reduction and processing including bias-subtraction, flat field correction, wavelength calibration and normalization were done using the Image Reduction and Analysis Facility \citep[\textsc{IRAF},][]{Tody1986}. For 5\,Peg, the OES spectra were of insufficient quality; therefore, high-quality data were retrieved from the public Melchiors archive \citep{Royer2024}, based on observations with the HERMES spectrograph on the Mercator telescope \citep{Raskin2014}.

\begin{figure}[H]
    \centering
    \includegraphics[width=0.8\linewidth]{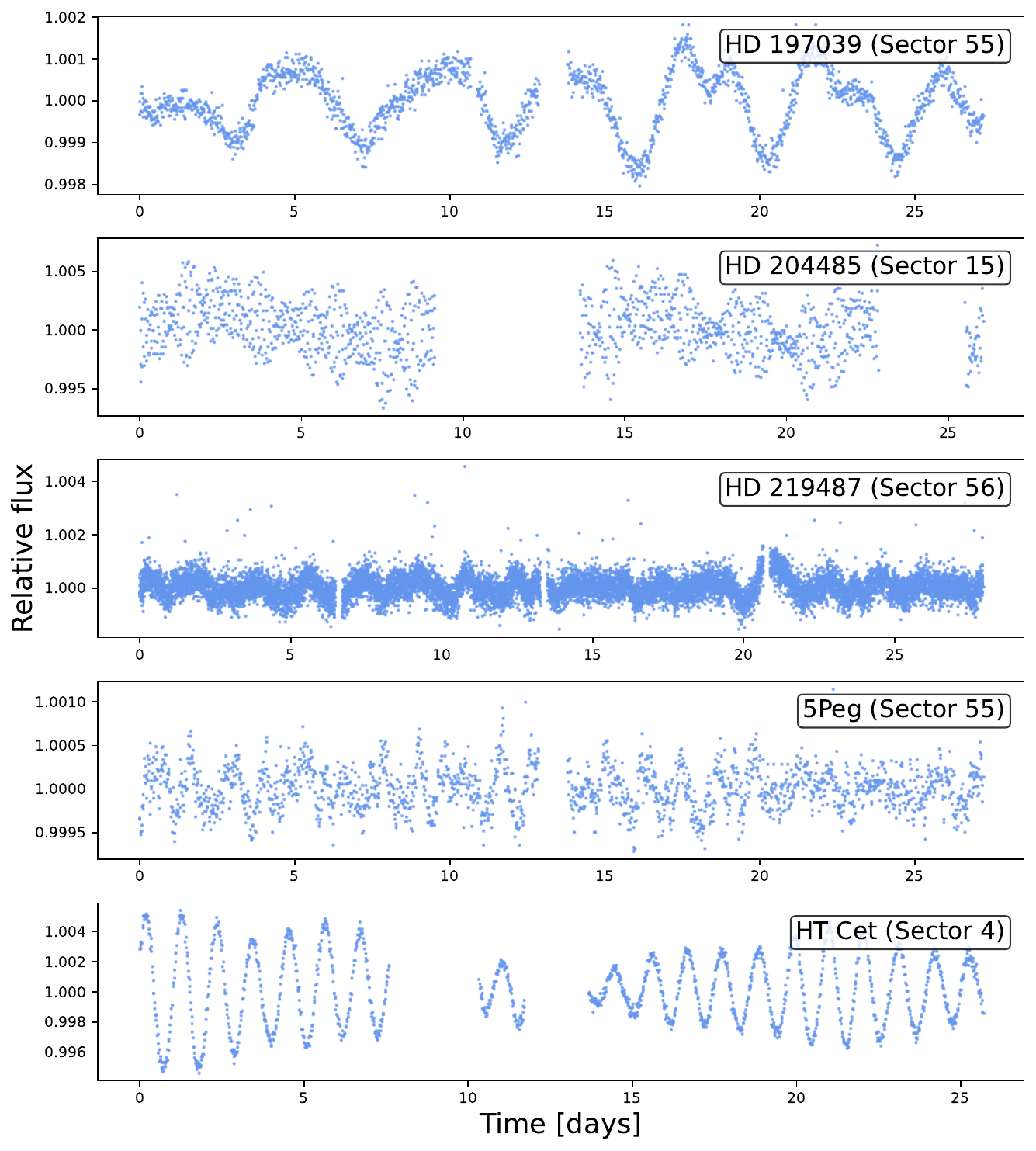}
    \caption{Light curves of selected objects from single sectors showing the basic features of the light variations. Note the different vertical scales.}
    \label{Fig:lightcurves}
\end{figure}

Photometric data were obtained from the TESS mission through the Science Processing Operations Center \citep[SPOC,][]{Jenkins2016,Caldwell2020}. We used Pre-search Data Conditioning Simple Aperture Photometry (PDCSAP) light curves with a 2-minute cadence. We downloaded four available sectors per star, providing sufficient temporal coverage for our analysis, by using \textsc{lightkurve} package \citep{Cardoso2018}. The lightcurves from a single sectors are shown in Fig.~\ref{Fig:lightcurves}. 

\section{Analysis}\label{Sect:Analysis}
\subsection{Spectra}

Spectral analysis was conducted via the \textsc{iSpec} framework \citep{Blanco-Cuaresma2014}. The process began with the re-normalisation of the spectra because \textsc{Iraf} normalisation was usually not of sufficient quality, and then cosmic ray artefacts were removed. Each spectrum was then corrected for radial velocity shifts by cross-correlating it with a synthetic mask based on atomic line data from the Vienna Atomic Line Database \citep[VALD,][]{Kupka1999}.

The wavelength range used for the cross-correlation analysis excluded regions contaminated by telluric lines. Balmer lines were not included in the fitting process due to their strong pressure broadening. Instead, we fitted as many Fe and metal lines as possible.

The initial atmospheric parameters — effective temperature ($T_{\rm eff}$ and surface gravity ($\log g$) — were adopted from the TESS Input Catalogue (TIC), accessed via the VizieR server \citep{Ochsenbein2000}. The microturbulent ($v_{\rm mic}$) and macroturbulent velocities ($v_{\rm mac}$) were estimated using the empirical relations given in \citet{Doyle2014}, \citet{Blanco-Cuaresma2019} and \citet{Worley2024}. The instrumental resolution was set to $R = 50000$ for the OES spectra and $R = 85000$ for the Hermes spectra \citep{Kabath2020,Raskin2011}.

The model atmospheres were based on ATLAS9 \citep{Castelli2003}, optimised for hotter stars. Spectral synthesis was performed using the MOOG \citep{Sneden2012} code, with solar abundances taken from \citet{Grevesse2007}. The adopted line list spanned 300–1100\,nm and was sourced from VALD. The alpha-enhancement and limb-darkening parameters were kept at their default values ($\alpha$ = 0.0, linear limb-darkening coefficient = 0.6). The parameters used are given in Tables \ref{Tab:Parameters} and \ref{Tab:vsini}.

\begin{figure}[htbp]
    \centering
    \begin{minipage}{1\textwidth}
        \centering
        \begin{subfigure}{\textwidth}
            \centering
            \includegraphics[width=0.98\textwidth]{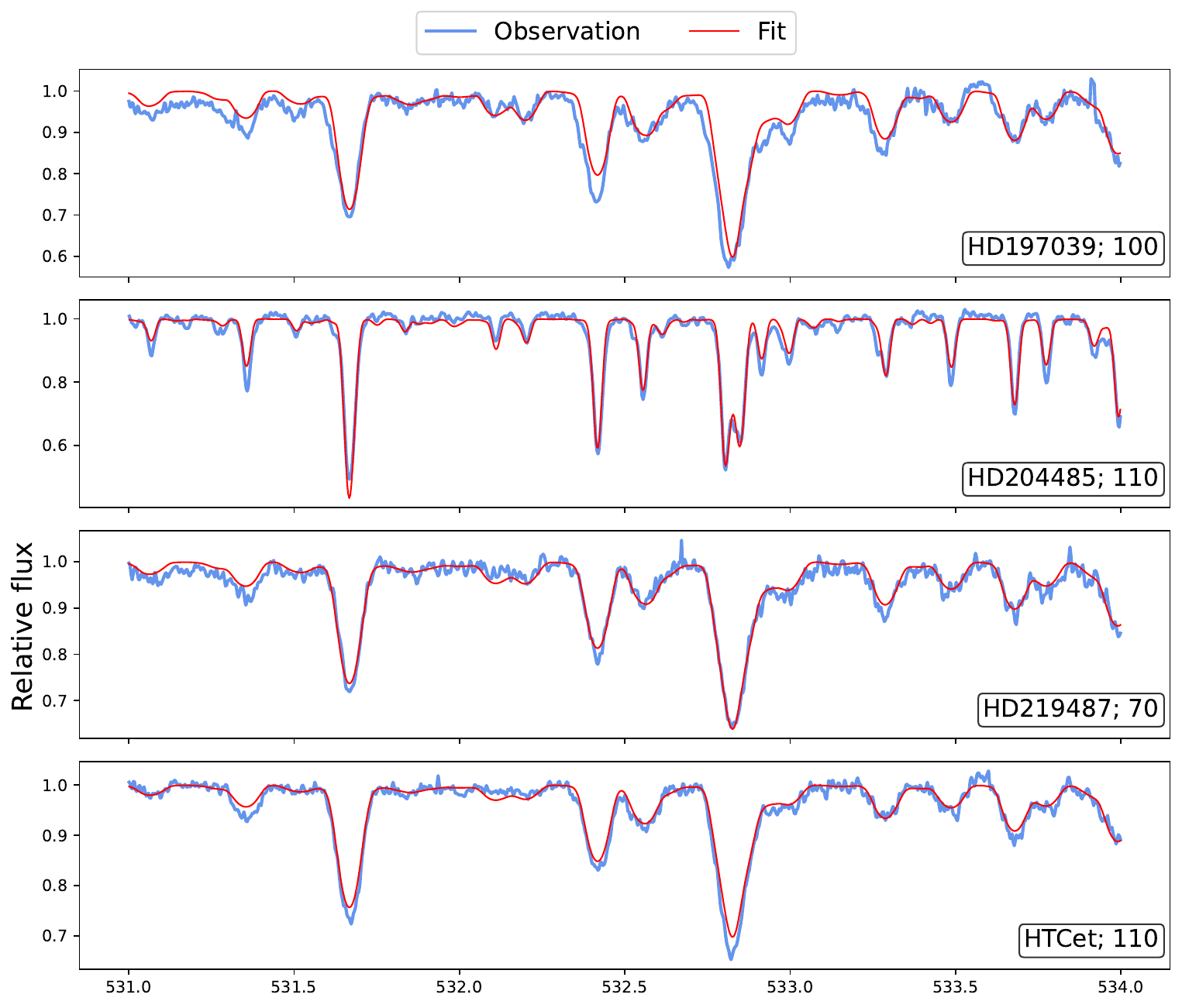}
        \end{subfigure}
        \vspace{0.1cm} 
        \begin{subfigure}{\textwidth}
            \centering
            \includegraphics[width=0.96\textwidth]{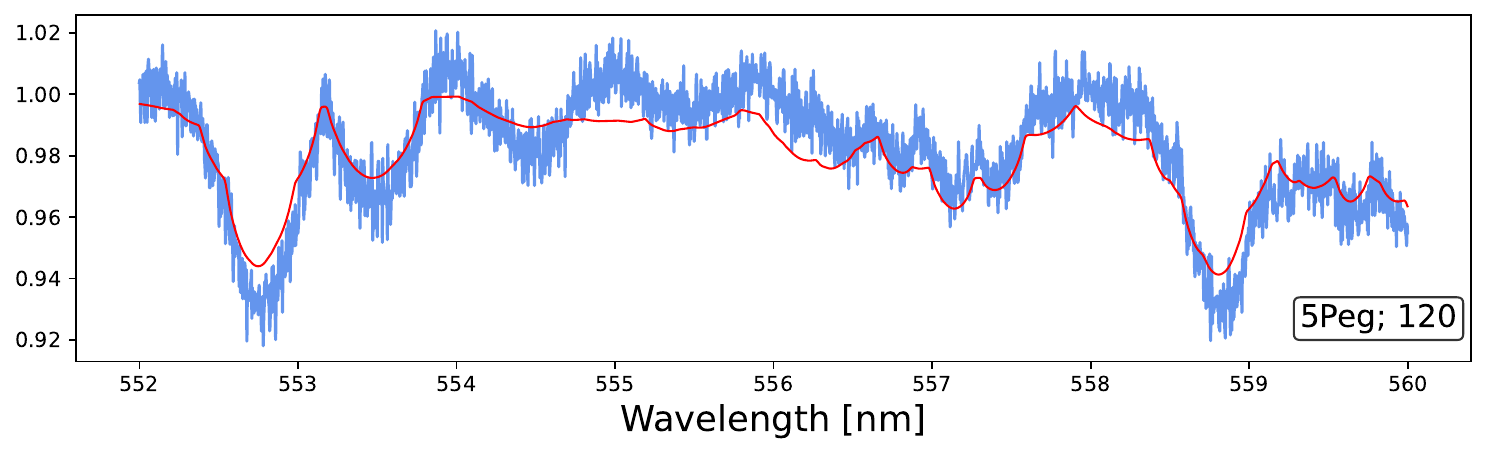}
        \end{subfigure}
    \end{minipage}
    \caption{The final fits of the spectra. Each panel is labeled with the star ID and its signal-to-noise ratio.}
    \label{Fig:Spectra}
\end{figure}

During fitting, the metallicity ([Fe/H]) and projected rotational velocity ($v \sin i$) were treated as free parameters. However, for HD\,197039, HD\,204485 and HD\,219487, allowing [Fe/H] to vary resulted in increased uncertainty in $v \sin i$ without improving the fit. In these cases, we assumed the metallicity to be zero and fixed it. For 5\,Peg we got [Fe/H]$=-0.44(2)$ and HT\,Cet -0.2(2). Each fit was iterated six times to ensure the convergence and stability of the solution. The final fits are shown in Fig.~\ref{Fig:Spectra}. The fits are reasonably good for most of the stars, with differences given mainly by the normalisation. In case of 5\,Peg, the fit is the worst but still reasonably good for the estimation of $v \sin i$.

The values of $v\sin i$ that we determined are in perfect agreement with the literature values (see Table~\ref{Tab:vsini}) showing a good reliability of our results.

\subsection{Photometry}

\begin{figure}[htbp!]
    \centering
    \includegraphics[width=0.7\linewidth]{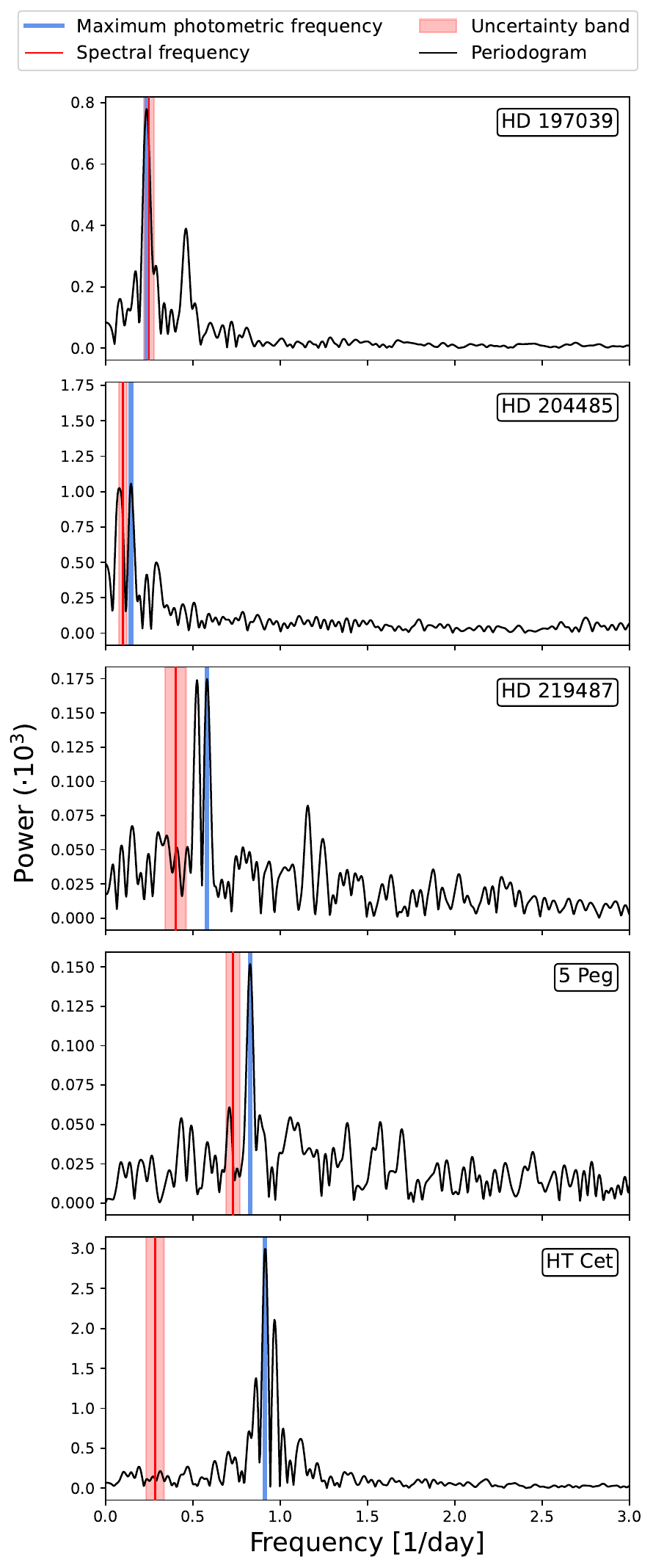}
    \caption{Periodogram with~the assumed spectroscopic rotational frequency $f_{\rm sp}$ (red) and the~brightness change frequency $f_{\rm ph}$ (blue).}
    \label{Fig:Periodogram}
\end{figure}

We processed photometric time series data using the \textsc{Lightkurve} Python package \citep{Cardoso2018} and computed Lomb–Scargle periodograms \citep{Lomb1976, Scargle1982} to identify the frequencies at which variability was most pronounced. We plotted the amplitude in relative flux units (Fig.~\ref{Fig:Periodogram}). 

Using the projected rotational velocities obtained from our spectroscopic analysis (Table~\ref{Tab:Parameters}) and the stellar radii from the TIC catalogue, we calculated the expected minimal rotational frequencies as 
\begin{equation}
    f_{\rm sp}=\frac{v\sin i}{2 \pi R}
\end{equation}
The values $f_{\rm sp}$ represent the lower limits of the rotational frequency because $v\sin i$ gives the bottom boundary of the real rotational velocity. By comparing the photometric and spectroscopic frequencies, we assessed whether the observed variability is consistent with rotational modulation or whether alternative mechanisms are more likely. The comparison of spectroscopic and photometric variability frequencies is shown in Fig.~\ref{Fig:Periodogram}.

\section{Discussion and conclusions}\label{Sect:Discussion}

This study examined whether the photometric variability observed in a sample of hot stars could be attributed to rotational modulation caused by surface spots. The projected rotational velocities derived from spectroscopy ($f_{\rm sp}$) were compared with the photometric frequencies ($f_{\rm ph}$) obtained from TESS data (see Table~\ref{Tab:vsini}). In HD\,197039, HD\,204485 and 5\,Peg, $f_{\rm sp}$ and $f_{\rm ph}$ are close to each other (see Fig.~\ref{Fig:Periodogram}). In addition, the frequency spectra of these three stars show well-defined peaks and, in the case of HD\,197039, HD\,204485, there are harmonics of the peaks suggesting rotational nature of the photometric variations due to surface inhomogeneities. 

In contrast, HT Cet — and possibly HD 219487 — exhibit complex frequency spectra indicative of $\gamma$ Doradus type pulsations \citep{Balona1994,Kaye1999}. However, we cannot rule out the possibility that some of the peaks originate from rotation. Assuming that the highest peaks in the frequency spectra reflect the real rotation frequencies, we can easily estimate the inclination of the rotation axes of the stars from the comparison of photometric and spectroscopic frequencies. The inclinations are shown in the last column of Table~\ref{Tab:Parameters}. From the proximity of $f_{\rm sp}$ and $f_{\rm ph}$ (see Table~\ref{Tab:vsini} and Fig.~\ref{Fig:Periodogram}) it is obvious that, under assumption of rotation, the inclinations of HD\,197039 and HD\,204485 are close to 90\,degrees. 

Although this method does not provide definitive proof of the rotational nature of the brightness variability, it can be used as a rough indicator. Our findings suggest that spot activity may extend to stars with effective temperatures in the range 6500--7200\,K, which is consistent with recent evidence of complex local magnetic fields generated in the (sub)surface layers of stars \citep{Henriksen2023,Antoci2025}. 

One of the sample stars, HD\,204485, was identified as a metallic chemically peculiar star by \citet{Maitzen1998}. Stars of this type (usually marked as CP1 or Am-Fm stars) show an underabundance of Ca and Sc and an overabundance of Fe-group and rare-earth elements \citep{Preston1974,Michaud1983}. Am-Fm stars are usually slowly rotating stars \citep{Abt1995}, which is the case of HD\,204485. In addition, this star appeared to be a $\delta$\,Sct type pulsator \citep{Breger2000} with the dominant frequency at 15.716\,c/d (see Fig.~\ref{Fig:lightcurves}). Pulsating stars of this type are quite common among Am-Fm stars \citep{Durfeldt-Pedros2024}. On the other hand, 5\,Peg should show $\delta$\,Sct pulsation \citep{Perry1987} but we did not detect any significant peak above 5\,c/d characteristic for this type of pulsations \citep{Uytterhoeven2011}. The remaining stars in our sample also do not show any additional variability.

Our results highlight the importance of combining photometric and spectroscopic data to identify spot candidates among hot stars. Further confirmation of the rotation and revelation of the nature of the spots would benefit from long-term spectroscopic monitoring enabling line-profile variation studies, Doppler imaging, and detailed chemical abundance analysis.

\acknowledgements
The results are based on a bachelor project at the Masaryk university in Brno, Czech Republic. MS acknowledges financial support of the Inter-transfer grant no LTT-20015. We acknowledge the usage of the data taken with the Perek telescope at the Astronomical Institute of CAS. 
The research of P.G. was supported by the Slovak Research and Development Agency under contract No. APVV-24-0160, and by internal grant No. VVGS-2023-2784 of the P. J. {\v S}af{\'a}rik University in Ko{\v s}ice and funded by the EU NextGenerationEU through the Recovery and Resilience Plan for Slovakia under project No. 09I03-03-V05-00008.

\bibliography{bibfile}

\end{document}